# Study of Cronin effect and nuclear modification of strange particles in d-Au and Au-Au collisions at 200 GeV in PHENIX

Dmitri Kotchetkov for the PHENIX collaboration[*]

*Department of Physics*
*University of California at Riverside, Riverside, CA 92521*

**Abstract.**
Effects of strangeness on nuclear modification in d-Au and Au-Au collisions at 200 GeV are studied, in order to quantify the effects of quark content and mass. Measurements of ratios of the yields in central collisions to the yields in peripheral collisions are performed for $\Lambda$ baryon and $\phi$ meson. Found results show little dependence of particle suppression or enhancement on mass and strange content, but rather prominent difference in nuclear modification between mesons and baryons.

From the analysis of hadrons produced in Au-Au collisions the phenomenon of suppression of hadron production in central collisions was discovered and reported [1,2]. Yet central d-Au collisions showed enhancement of hadrons [3], as expected from the Cronin effect [4]. However, there are differences in the behavior for protons and pions. It was found, that in central Au-Au collisions protons are not suppressed, when their transverse momenta are larger than 2 GeV/c, though pions show suppression within measured transverse momentum range (up to 10 GeV/c for $\pi^0$). Also, in central d-Au collisions observed nuclear enhancement is larger for protons. Here we expand the study of nuclear modification to include strange mesons and baryons.

To study nuclear modification in Au-Au and d-Au collisions it is convenient to introduce a ratio of the particle yield in central collisions to the particle yield in peripheral collisions $R_{CP}$. Both yields are normalized by corresponding numbers of binary collisions:

$$R_{CP} = \frac{Yield(central)/<N_{COLL}(central)>}{Yield(peripheral)/<N_{COLL}(peripheral)>}$$

Then, the following questions can be addressed: 1) how the presence of strange quarks influences $R_{cp}$; 2) how the number of strange quarks affects nuclear modification, i.e. is there a difference between strange baryons and strange mesons; 3) is $R_{cp}$ a function of mass of the strange particle. Measuring $R_{cp}$ parameters for $\Lambda$ baryon and $\phi$ meson is the first step in answering these questions.

PHENIX detector has the capability to measure the properties of strange particles through the reconstruction of their decay products. Both $\Lambda$ and $\phi$ can be reconstructed in the PHENIX Central Spectrometer [5]. It has two opposing arms, each covering $90^0$ of azimuthal angle and 0.7 units of pseudorapidity. Beam-Beam Counters define a collision vertex and the reference for measuring hadron time of flight. Drift and Pad Chambers

---

[*] For the full PHENIX Collaboration author list, see Appendix "Collaborations" of this volume

reconstruct charged tracks. Two different detectors are employed to find a time of flight of the charged hadron. In one arm of the Central Spectrometer there is a Time of Flight Hodoscope (TOF) with $45^0$ of azimuthal acceptance and 115 ps of time of flight resolution. It is completed by two sectors of a lead-scintillator Electromagnetic Calorimeter (EMCal), each covering $22.5^0$ of azimuthal acceptance, having time of flight resolution equal to 700 ps. The opposite arm of the Central Spectrometer has 4 similar sectors of Electromagnetic Calorimeter. The radial distance from the beam line to each sector of the EMCal and TOF is approximately 5.1 m.

The decay of Λ baryon into proton-pion pair is highly asymmetric. Two analysis procedures are used to estimate the systematic errors. First, pions and protons are identified either in EMCal or TOF, then all possible combinations of p-π pairs from same collision event build the Λ invariant mass spectrum. The combinatorial background is formed by combining protons and pions from mixed events.

The second approach takes into account asymmetry of Λ decay. Since the Time of Flight Hodoscope has superior time resolution compared to EMCal, one can identify protons only in TOF (with good particle identification up to 3 GeV of transverse momentum) combined with any hadron detected either in TOF or EMCal. A total of 63 million d-Au collision events were used to extract yields of Λ baryon. Particle yields were used to find $R_{cp}$ of Λ for four different centrality classes, 0-20%, 20-40%, 40-60%, and 60-88.5%, as determined by the multiplicity in the South (Au-side) Beam-Beam Counter.

The ϕ meson is reconstructed in the PHENIX detector via K+K- decay channel. Kaons can be identified either in EMCal or TOF. Similar mixing technique is employed to build a combinatorial background. Yields of ϕ meson are found from the analysis of 19 million Au-Au collision events in 0-10%, 10-40%, and 40-92% centrality classes.

To look at Cronin enhancement of Λ in d-Au collisions we plot $R_{cp}$ factor of Λ baryons together with $R_{cp}$ of protons, pions and kaons. Figures 1-3 show $R_{cp}$ factors, when yields from central collisions are compared to the particle yields in events with centralities equal to 60-88.5%. Figure 1 represents the case, when collisions with 0-20% centrality are considered to be central. There is a noticeable increase of the particle yields in central collisions. Figure 2 represents the results, when events with centralities of 20-40% are taken to be central. Figure 3 shows the class of events with 40-60%. The enhancement decreases with decreasing centrality. The enhancement of lambda, as a strange baryon, tends to replicate the enhancement of proton. Note that the mass of lambda is close to that of proton. There is no visible effect of strangeness on Cronin enhancement.

Results of nuclear modification of ϕ meson in central Au-Au collisions are shown in Figure 4. $R_{cp}$ of ϕ meson is computed as a ratio of the yield in the collisions with centrality of 0-10% to the yield in the collisions with centrality of 40-92%. In the same figure, ratios of yields of protons and neutral pions in 0-10% central collisions to yields in 60-92% central collisions are plotted. It is seen that ϕ remains suppressed for all measured transverse momenta, similar to the pion, even though the value of mass of ϕ meson is close to that of proton.

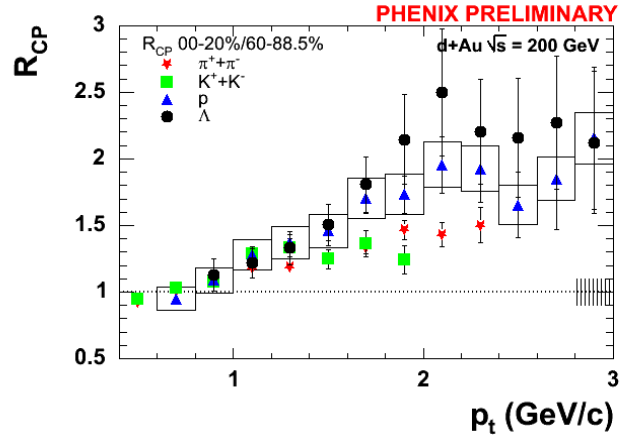

Figure 1. $R_{cp}$ of identified hadrons (0-20% d-Au central collisions).

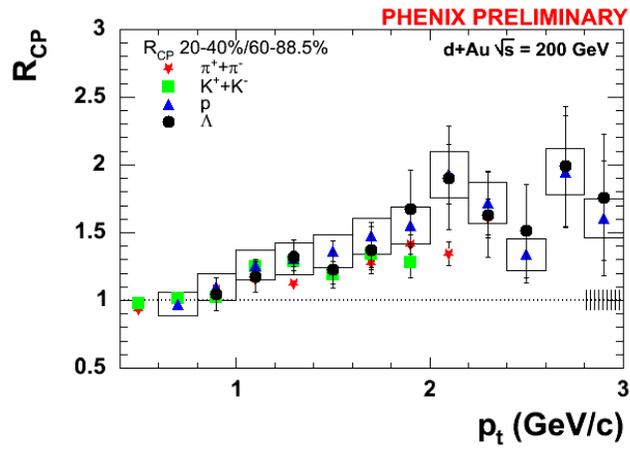

Figure 2. $R_{cp}$ of identified hadrons (20-40% d-Au central collisions).

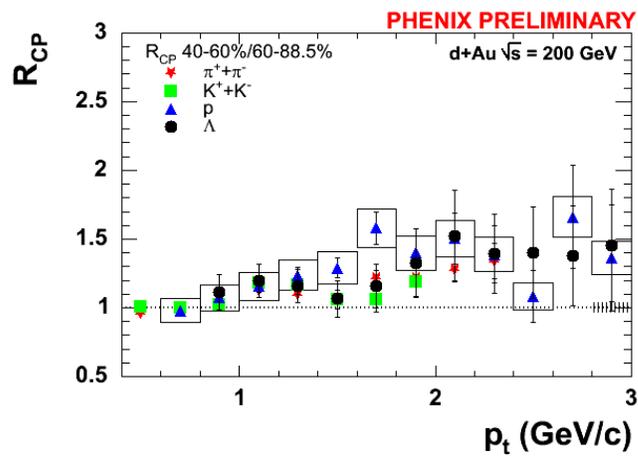

Figure 3. $R_{cp}$ of identified hadrons (40-60% d-Au central collisions).

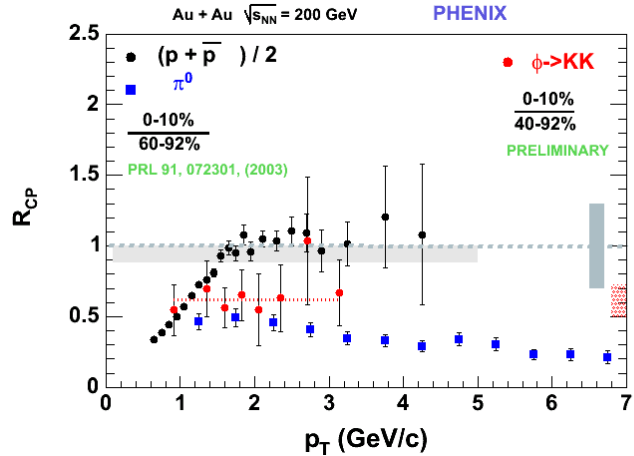

Figure 4. $R_{cp}$ of φ meson, protons and neutral pion (0-10% Au-Au central collisions).

Statistical errors were calculated from particle yields in each transverse momentum bin. Binary collisions scaling errors are in the order of 7-10%. Systematic errors were found through varying track matching, particle identification, background subtraction and other requirements, while extracting particle yields. These errors are less than 10%.

Comparison of $R_{cp}$ values of Λ and proton in d-Au collisions might hint that strangeness has little, if any, effect on nuclear modification. Current $R_{cp}$ results of φ meson compared with $R_{cp}$ results for proton in central Au-Au collisions show indications supporting this conclusion, but require improved statistics to reduce rather large error bars from φ meson measurements. Also, suppression of φ, and absence of suppression of proton show that nuclear modification is a weak function of particle mass. From absence of suppression of Λ baryon in central Au-Au collisions reported by STAR collaboration [6] and suppression of φ meson in central Au-Au collisions reported by PHENIX collaboration one can conclude that number of quarks might influence nuclear modification, so there is a difference in evolution of meson and baryon matters. A theory of quark recombination might be a suitable model for this phenomenon [7].